%Paper: hep-th/9305037
%From: Philippe Di Francesco <philippe@amoco.saclay.cea.fr>
%Date: Mon, 10 May 1993 05:19:30 +0000

%
%
%
%05-10-93
%
%         Ramond Sector Characters and N=2 L.G. Models
%
%           by P. Di Francesco and S. Yankielowicz
%
%
%
%

\input harvmac

\overfullrule=0mm

\def\encadre#1{\vbox{\hrule\hbox{\vrule\kern8pt\vbox{\kern8pt#1\kern8pt}
\kern8pt\vrule}\hrule}}
\def\encadremath#1{\vbox{\hrule\hbox{\vrule\kern8pt\vbox{\kern8pt
\hbox{$\displaystyle #1$}\kern8pt}
\kern8pt\vrule}\hrule}}

%Macros
%%%%%%%%%%%%%%%%%%%%%%%%%%%%%%%%%%%%%%%%%%%%%%%%%%%%%%%%%%%%%%%
%%%%%%%%%%%%%%%%%%%DEFINITIONS%%%%%%%%%%%%%%%%%%%%%%%%%%%%%%%%%
%
\def\frac#1#2{\scriptstyle{#1 \over #2}}

%
%%%%%%%%%%%%%%%%%%%CALIGRAPHIC LETTERS%%%%%%%%%%%%%%%%%%%%%%%%%
%

\def\CV{{\cal V}}		

\def\({ \left( }
\def\){ \right) }
%%%%%%%%%%%%%%%%%%%%MATH CHARACTERS%%%%%%%%%%%%%%%%%%%%%%%%%%%%
%

%\font\numbers=cmu10 scaled\magstep1

\def\IR{\relax{\rm I\kern-.18em R}}
\font\cmss=cmss10 \font\cmsss=cmss10 at 7pt
\def\IZ{\relax\ifmmode\mathchoice
{\hbox{\cmss Z\kern-.4em Z}}{\hbox{\cmss Z\kern-.4em Z}}
{\lower.9pt\hbox{\cmsss Z\kern-.4em Z}}
{\lower1.2pt\hbox{\cmsss Z\kern-.4em Z}}\else{\cmss Z\kern-.4em Z}\fi}
\def\inbar{\,\vrule height1.5ex width.4pt depth0pt}
\def\IB{\relax{\rm I\kern-.18em B}}
\def\IC{\relax\hbox{$\inbar\kern-.3em{\rm C}$}}
\def\ID{\relax{\rm I\kern-.18em D}}
\def\IE{\relax{\rm I\kern-.18em E}}
\def\IF{\relax{\rm I\kern-.18em F}}
\def\IG{\relax\hbox{$\inbar\kern-.3em{\rm G}$}}
\def\IH{\relax{\rm I\kern-.18em H}}
\def\II{\relax{\rm I\kern-.18em I}}
\def\IK{\relax{\rm I\kern-.18em K}}
\def\IL{\relax{\rm I\kern-.18em L}}
\def\IM{\relax{\rm I\kern-.18em M}}
\def\IN{\relax{\rm I\kern-.18em N}}
\def\IO{\relax\hbox{$\inbar\kern-.3em{\rm O}$}}
\def\IP{\relax{\rm I\kern-.18em P}}
\def\IQ{\relax\hbox{$\inbar\kern-.3em{\rm Q}$}}
\def\IGa{\relax\hbox{${\rm I}\kern-.18em\Gamma$}}
\def\IPi{\relax\hbox{${\rm I}\kern-.18em\Pi$}}
\def\ITh{\relax\hbox{$\inbar\kern-.3em\Theta$}}
\def\IOm{\relax\hbox{$\inbar\kern-3.00pt\Omega$}}

%%%%%%%%%%%%%%%%%%%%%%%%%%%%%%%%%%%%%%%%%%%%%%%%%%%%%%%%%%%%%%%%%

%Mes Macros

%\def\Z{_N Z}

%%%%%%%%%%%%%%%%%%%%Greek letters%%%%%%%%%%%%%%%%%%%%%%%%%%%%%%%%%%

\def\Gl{\lambda}

\def\GS{\Sigma}

%%%%%%%%%%%%%%%%

%
%
%
%
%
\Title{SPhT 93/049--TAUP 2047-93\ \ hep-th/9305037}
{{\vbox {\vskip-.5in
\centerline{Ramond Sector Characters}
\bigskip
\centerline{and N=2 Landau--Ginzburg Models}
}}}
\centerline{P. Di Francesco,}
\medskip\centerline
{\it Service de Physique Th\'eorique de Saclay
\footnote{$^{\#}$}{
Laboratoire de la Direction des Sciences et de la Mati\`ere
du Commissariat \`a l'Energie Atomique},}
\centerline{\it 91191 Gif sur Yvette Cedex, France,}
\medskip
\centerline{and}
\medskip
\centerline{S. Yankielowicz \footnote{$^{\$}$}{Work supported
in part by the US--Israel Binational Science Foundation
and the Israel Academy of Sciences.},}
\medskip\centerline{\it School of Physics and Astronomy,}
\centerline{\it Beverly and Raymond Sackler Faculty of Exact
Sciences,}
\centerline{\it Tel--Aviv University,
Ramat--Aviv, Tel--Aviv 69978, Israel.}
\vskip .2in
We give a direct proof of the new "product" expression
for the Ramond sector characters of N=2 minimal
models recently suggested by E. Witten. Our construction
allows us to generalize these expressions to the D and E
series of N=2 minimal models, as well as to other N=2
Kazama--Suzuki coset models such as
$SU(N)\times SO(2(N-1))/SU(N-1)\times U(1)$.
We verify that these expressions
indeed coincide with the corresponding Landau--Ginzburg
"elliptic genus", a certain topologically invariant twisted
path integral with the effective Landau--Ginzburg action,
which we obtain by using Witten's method.
We indicate how our approach may be used to construct
(or rule out) possible Landau--Ginzburg potentials for
describing other
N=2 superconformal theories.

\Date{05/93}
%\draftmode
%
\newsec{Introduction}

N=2 superconformal theories have been under active
investigation during recent years.
They are of interest for the construction of vacuum states
of super-string theories as well as for the investigation
of topological theories.
The structure of these theories is very much constrained
by the N=2 algebra and its representations.
This applies in particular to the chiral fields in
the Neveu-Schwarz (NS) sector which form a ring under
the fusion rules
\ref\RING{D. Gepner, Nucl. Phys. {\bf B322} (1989) 65.}\
\ref\EMLG{D. Gepner, Comm. Math. Phys. {\bf 141} (1991) 381.}.
By spectral flow these fields are in one to one correspondence
with the ground states in the Ramond (R) sector
\ref\SPEC{A. Schwimmer and N. Seiberg, Phys. Lett. {\bf B184}
(1987) 381;
D. Gepner, "Lectures on N=2 String Theory",
Proceedings of the Trieste Spring School 1989.}.
Moreover, the chiral primary fields are the ones which survive
in the topological theory obtained from
the N=2 theory upon twisting by the U(1) current
\ref\CHIR{E. Witten, Comm. Math. Phys. {\bf 118} (1988) 411;
T. Eguchi and S.-K. Yang, Mod. Phys. Lett. {\bf A5} (1990) 1693.}.

  Another realization of N=2 supersymmetric theories is in terms
of N=2 Landau--Ginzburg (LG) models
\ref\ACT{C. Vafa and N. Warner, Phys. Lett. {\bf B218} (1989) 51;
E. Martinec, Phys. Lett. {\bf B217} (1989) 431;
{\it Criticality, Catastrophes
and Compactifications}, Physics and Mathematics of Strings,
ed. L. Brink, D. Friedan and A. M. Polyakov (World Scientific, 1990).}.
The LG theory is given by an action
$$    S= \int{d^2z d^4\theta \ K([\Phi_i,{\bar \Phi}_i])}+
\int d^2z d^2\theta \
W([\Phi_i]) +c.c.$$
where the $\Phi_i(z,\theta)$ (${\bar \Phi}_i({\bar z},{\bar \theta}$)
are a set of chiral ( anti--chiral) N=2 superfields.
$K([\Phi_i,{\bar \Phi}_i]$) is the kinetic term and
$W([\Phi_i])$ is the superpotential.
Given a potential one can define a ring of
polynomials in the $\Phi_i$ modulo the LG equation of
motion dW=0 \ACT\ \EMLG, i.e.
$$ {\cal R} = \IC[\Phi_i]/dW \qquad      i=1,....,M   $$

 At the conformal point, the LG superpotential
should be a quasi-homogeneous function so as to allow for a grading of
the superfields. This grading is related to the so called
R--symmetry of the N=2 LG theory. As we perturb away from the
critical point both the superpotential and the fields get modified,
however the structure of the perturbed chiral ring is still
described by $\cal R$. The basic
conjecture is that at the UV conformal point the behaviour
of the theory is governed by the superconformal N=2 RCFT whose chiral
ring is isomorphic to the ring $\cal R$.
There is a one to one correspondence  between the chiral fields of the
LG theory and the chiral fields of the underlying N=2 RCFT.
Moreover, the
grading is associated with the U(1) charge of the N=2 algebra.
{}From this point
of view any N=2 LG theory corresponds to a N=2 RCFT.
It is still an open question to better understand and classify all
N=2 RCFT's which admit a LG
description.

  There is by now a bulk of "circumstancial" evidence which support this
conjecture and make it very plausible.
This involves approximate RG flows
\ref\GW{P.S. Howe and P.C. West, Phys. Lett. {\bf B227} (1989) 397.}.
Moreover, the superpotential encodes a lot of
information concerning the underlying N=2 RCFT.
{}From the grading one can
determine the U(1) charges and the corresponding central charge
\ref\CV{C. Vafa and N. Warner, Phys. Lett. {\bf B218} (1989) 51.}.
The chiral ring itself is isomorphic to the polynomial
ring associated with the LG potential, and this is reflected in
the properties of the chiral fields \EMLG\
\ref\CHF{ W. Lerche, C. Vafa and N. Warner, Nucl. Phys.
{\bf B324} (1989) 427;
S. Cecotti, L. Girardello and A. Pasquinucci, Nucl. Phys. {\bf B238}
(1989) 701 and Int. J. Mod. Phys. {\bf A6} (1991) 2427;
S. Cecotti, Int. J. Mod. Phys. {\bf A6} (1991) 1749.}.
Recently an important step toward proving the conjecture has been
made by Witten
\ref\WIT{E. Witten,
{\it On the Landau-Ginzburg description of
N=2 minimal models,} preprint IASSNS--HEP--93/10 (1993).}\
within the framework of
the A series of N=2 minimal models.
Assuming that the conjecture holds, Witten related
certain characters in the
Ramond sector of the N=2 RCFT to the elliptic
genus of the corresponding LG
theory.
This elliptic genus is given by the path integral with certain twisted
boundary conditions and can be effectively computed in the LG theory.
As noted
by Witten, an important feature of the elliptic genus is the fact that for
supersymmetric models it remains conformally invariant even if the model
itself is not.
Thus, it remains the same as we approach the UV limit.
This way
an interesting "product" formula for these particular
Ramond sector characters
was obtained. The formula was checked by expanding it
to few low orders in q
and comparing to the known character formulas
\ref\CHFOR{ V. K. Dobrev, Phys. Lett. {\bf B186} (1987) 43;
E. Kiritsis, Int. J. Mod. Phys. {\bf A3} (1988) 1871;
F. Ravanini and S.-K. Yang, Phys. Lett. {\bf B195} (1987) 202;
Z. Qiu, Phys. Lett. {\bf B198} (1987) 497;
J. Distler and Z. Qiu, Nucl. Phys. {\bf B336} (1990) 533;
T. Eguchi, T. Kawai, S. Mizoguchi and S.-K. Yang,
Rev. Mod. Phys. {\bf 4} (1992) 329;
R. Kedem, T. R. Klassen, B. m. McCoy and E. Melzer,
{\it Fermionic Sum Representations for Conformal Field
Theory Characters}, Stonybrook preprint (1993).}.

  In section 2 of the paper we give a direct proof
of this character formula and generalize it to the D and E
modular invariants of the N=2 minimal models
based on the coset $SU(2)_k/U(1)$.
It is based on a mathematical lemma on elliptic
modular functions which we prove in appendix A.
In section 3 we
generalize it to
Kazama--Suzuki N=2 theories
\ref\KS{Y. Kazama and H. Suzuki, Nucl. Phys. {\bf B321} (1989) 232.}\
based on the coset
$SU(N)_k\times SO(2(N-1))_1/SU(N-1)_{k+1} \times U(1)$.
{}From our
analysis it will be clear that it is the U(1) charge of the N=2 algebra which
plays the crucial role, almost determining
the whole structure.
The elliptic genus in both the superconformal and the LG frameworks
is strongly constrained by the transformations of
the various fields under the U(1) symmetry.
For the N=2 superconformal theory it is the U(1) charge of
the Ramond sector while in the LG approach it is the
R--symmetry charge of the chiral superfields.
In section 4 we discuss and emphasize this aspect of our
approach.
We also address the question of the uniqueness of the
identification of the LG
potential and the "stability" of the elliptic genus
under "massive" perturbations.
The important lesson which we would like to convey is that the
identification between the N=2 superconformal theory and
the corresponding LG
model crucially depends on the U(1) grading.
We shall give examples to clarify
this point and discuss the constraints it imposes for the
existence of a  LG description. It is clear that the knowledge
of the elliptic genus either gives some information about the
LG potential, or rules out its existence. In the first occurence,
one may have a new tool for hunting LG potentials; in the latter,
we get a criterium for non--existence of Landau--Ginzburg potential
(at least with the grading of fields matching that of the chiral
Ramond states), namely that the elliptic genus cannot be put
into a nice product form.
This last point will be illustrated with simple examples.

\newsec{The SU(2) case.}
\subsec{Landau--Ginzburg description of N=2 superconformal $SU(2)_k$
theories.}

In a recent paper \WIT,
E. Witten proposed a link between the N=2 superconformal minimal
theories based on SU(2) and their effective description in terms
of a N=2 Landau--Ginzburg superfield.  In both theories, he computed the
so--called elliptic genus of the theory, a particular toro\" \i dal
twisted partition
function. On one hand it is a particular linear combination of
characters for Ramond states, on the other hand it can be directly
computed within the Landau--Ginzburg framework \WIT.
More precisely, this function is defined as
\eqn\ellipgen{ Z_2(u|\tau)={\sum_{l}}'
{\rm Tr}_{{\cal R}_{l}} (-1)^{F_L} q^{H_L}
e^{i \gamma_L J_{0,L}}, }
where $F_L$, $H_L$, $J_{0,L}$ and $\gamma_L$ denote respectively
the fermion number, the hamiltonian
$H_L=L_0-{c \over 24}$, the U(1) symmetry generator zero component
and associated charge of the left--moving Ramond states,
and the sum extends over the states with
vanishing right--moving hamiltonian $H_R=0$ and $U(1)$ charge
$\gamma_R=0$. In eqn.\ellipgen, ${\cal R}_{l}$ denote the
Ramond sector representation of the N=2 superalgebra containing
a ground state of $H_L=0$, and we denote by $u={\gamma_L \over
2 \pi (k+2)}$ and $q=e^{2i \pi \tau}$.

Such representations are well known in the context of N=2 minimal
superconformal theories based on SU(2), and correspond to the Ramond
sector analogues of the Neveu--Schwarz chiral fields, obtained
from those by the standard
spectral flow \SPEC.
The corresponding characters are obtained by considering the
$N=2$ theory as a ${SU(2)_k \over U(1)} \times U(1)$
coset, and they read
\ref\GCH{Z. Qiu, Phys. Lett. {\bf B198} (1987) 497;
D. Gepner, Nucl. Phys. {\bf B296} (1988) 757.}
\eqn\chartwo{ \chi_{l}^{l}(z|\tau) =\sum_{j \  {\rm mod} \  k}
C_{l+4j}^{l}(\tau) \ \Theta_{ (k+2)(4j-1)+2(l+1);2k(k+2)} (z|\tau) }
where $C_m^l(\tau)$, $|m| \leq l \leq k$,
denote the parafermionic string functions
and the U(1) theta function is defined as
\eqn\defthet{ \Theta_{m,p}(x|\tau)=\sum_{n \in \IZ} q^{p(n+m/2p)^2}
e^{4i \pi p x (n+m/2p)} }
The elliptic genus for the minimal N=2 superconformal theory
is just the sum of the above characters
\eqn\genelip{ K_2(z|\tau)=\sum_{l=0}^k \chi_l^l(z|\tau).}

It is believed that there exists an effective description
of the N=2, A type (referring to the fact that all fields
$l=0,..,k$ are present in the theory), superconformal theory in
terms of a N=2 superfield $\Phi$, governed by the action
$$ S= \int d^2x d^4 \theta \ \Phi {\bar \Phi}
+\{ \int d^2x d^2 \theta \ {{\Phi}^{k+2} \over k+2} + {\rm c.c.}\}.$$
The direct Landau--Ginzburg computation is made possible by the
following argument \WIT: the elliptic
genus \ellipgen\ is a topological
invariant, therefore independent on an overall
arbitrary scaling parameter $\epsilon$ multiplying the potential.
Moreover the $\epsilon \to 0$ limit is regularized by the twist imposed
on the various fields of the theory. Hence one can take $\epsilon=0$
and perform a simple free field computation.
Moreover, from a careful study of the symmetries of
the potential, one gets the U(1) transformations
of the bosonic lower component and the fermionic components of the
superfield
$\Phi=\phi+\theta_+ \psi^{+} + \theta_{-} \psi^{-}
+\theta_+ \theta_- F$ after the standard
gaussian integration
over the upper component $F$, namely
\eqn\uonetwo{\eqalign{
\phi &\to e^{2 i \pi u} \phi \cr
\psi^{+} &\to e^{2i \pi u} \psi^{+} \cr
\psi^{-} &\to e^{-2i \pi (k+1)u} \psi^{-} \cr}}
and conjugate transformations for the conjugate $\bar \Phi$ components.
Putting together the contributions to \ellipgen\ of all the modes of
the left and right movers, Witten obtains a simple product formula
\eqn\witres{ Z_2(u|\tau) =e^{-i\pi u k}
\prod_{n\geq 0} { (1-q^n e^{2 i \pi (k+1)u}) \over (1-q^n e^{2i\pi u})}
\prod_{n \geq 1}{ (1-q^n e^{-2i \pi (k+1) u}) \over
(1-q^n e^{-2i\pi u})} }
This can be recast in terms of the first Jacobi theta function
\eqn\deftet{\Theta_1(u|\tau)
=-i e^{-i\pi u}\prod_{n \geq 0} (1-q^n e^{2i\pi u})
\prod_{n \geq 1} (1-q^n)(1-q^n e^{-2i\pi u}).}
We find
\eqn\witrevis{ Z_2(u|\tau)={ \Theta_1((k+1)u|\tau) \over
\Theta_1(u|\tau)}. }
A first step toward the identification of the N=2 superconformal
theory based on SU(2) and the Landau--Ginzburg theory
of the N=2 superfield $\Phi$ is the identification of elliptic
genera \genelip\ and \witrevis, which amounts to
\eqn\witconj{Z_2(u=2z|\tau)=K_2(z|\tau).}

\subsec{The proof of Witten's character formula expressing
the elliptic genus in the A type SU(2) case.}

We wish now to prove the identity between \genelip\ and \witrevis.
The proof goes in three steps: first we compute the behaviour
of the elliptic genus of the N=2 superconformal theory expressed as
\genelip\ under the transformations $z \to z+1$ and $z \to z+\tau$.
Comparing it to that of the Landau--Ginzburg expression \witrevis,
we find that the ratio $K_2/Z_2$ is an elliptic function of $z$
(i.e. $1$ and $\tau$--periodic).
The second step uses the modular covariance of both
versions of the elliptic genus to prove that $K_2/Z_2$ is in
addition a modular form of weight zero.
The last step uses standard
elliptic function theory and the $q\to 0$ limit of the ratio to
conclude that it is a constant, which turns out to be $1$.

By a straightforward use of equations \chartwo\--\genelip,
we find that
\eqn\ktwoonetau{\eqalign{
K_2(z+1|\tau)&= K_2(z|\tau) \cr
K_2(z+\tau|\tau)&= e^{-4i \pi k(k+2)(\tau+2z)} K_2(z|\tau). \cr}}
On the other hand, using the transformations of the Jacobi theta
function
\eqn\thettrans{\eqalign{
\Theta_1(u+n|\tau)&=(-1)^n \Theta_1(u|\tau) \cr
\Theta_1(u+n\tau|\tau)&= (-1)^n e^{-i\pi n(n\tau+2u)}
\Theta_1(u|\tau), \cr }}
it is easy to see that
\eqn\elitran{\eqalign{
Z_2(2(z+1)|\tau)&=Z_2(2z|\tau)\cr
Z_2(2(z+\tau)|\tau)&=e^{-4i\pi k(k+2)(\tau+2z)}Z_2(2z|\tau) \cr}}
Therefore the ratio $K_2(z|\tau) /Z_2(2z|\tau)$ is an elliptic function
of $z$ with periods $1$ and $\tau$, and has a finite number of single
poles due to zeroes of the denominator. Standard elliptic function
theory enables to write
\ref\WWAT{E. Whittacker and G. Watson, {\it Modern Analysis},
Cambridge University Press (1958).}\
$$ {K_2(z|\tau) \over Z_2(2z|\tau)} = A \prod_{i=1}^p
{\Theta_1(z-a_i|\tau) \over \Theta_1(z-b_i|\tau)}$$
where $a_i(\tau)$, resp. $b_i(\tau)$
denote the zeroes and poles of the elliptic
function on its fundamental domain.

The second step of the proof uses the modular covariance of $K_2$
and $Z_2$. On one hand, from the modular transformations of the
characters
\CHFOR\
$$\eqalign{
C_m^l(-{1 \over \tau})&=(-i\tau k(k+2))^{-1/2}
\sum_{l',m'} \sin\pi {(l+1)(l'+1)\over
k+2} e^{-i\pi {m m' \over k}} C_{m'}^{l'}(\tau) \cr
\Theta_{m,p}({z \over \tau}| -{1 \over \tau})&=
({-i\tau \over 2p})^{1/2} e^{2i\pi p {z^2 \over \tau}}
\sum_{m'} e^{-i\pi {mm' \over p}} \Theta_{m',p}(z|\tau), \cr}$$
we find that
$$K_2({z \over \tau}|-{1 \over \tau})=e^{4i \pi k(k+2){z^2 \over \tau}}
K_2(z|\tau).$$
On the other hand, using the standard modular transformation of the
Jacobi theta function
$$\Theta_1({z \over \tau}|-{1 \over \tau})=i (-i\tau)^{1/2}
e^{i\pi {z^2 \over \tau}} \Theta_1(z|\tau),$$
we get
$$Z_2({2 z \over \tau}|-{1 \over \tau})=e^{4i \pi k(k+2){z^2 \over \tau}}
Z_2(2z|\tau),$$
so that the ratio $K_2(z|\tau)/Z_2(2z|\tau)$ is invariant under the
modular "S" transformation $(z,\tau) \to ({z \over \tau},-{1 \over \tau})$.
Finally, it is easy to see that both expressions for the
elliptic genus are invariant under the "T" transformation
$\tau \to \tau + 1$: for $K_2$, it is a direct consequence of the
choice of Ramond states with $L_0=c/24$ (the characters are
transformed under T by a phase factor $\exp(2i\pi (h-c/24)=1$
here\foot{Note that
this is a general, built--in property of the elliptic genus.
It will apply to all the other cases we will consider.
We suspect also that the "S"--covariance is a generic property
of elliptic genera, once it is understood as some
two point correlator of "twist" operators.});
for $Z_2$, it is a consequence of the Jacobi theta function
transformation
$\Theta_1(z|\tau+1)=e^{i\pi /4} \Theta_1(z|\tau)$.

The last step uses the
$\tau \to i \infty$ (or $q \to 0$)
limit of the elliptic function. Let us now  prove that
\eqn\sipro{\lim_{\tau \to i \infty} {K_2(z|\tau) \over Z_2(2z|\tau)}=1,}
as a consequence of
the limits of $Z_2$ and $K_2$.
We have:
\eqn\limztwo{ \lim_{q \to 0} Z_2(2z|\tau)={\sin 2\pi(k+1)z \over
\sin 2 \pi z} }
and the contribution to the limit of $K_2$ only involves the U(1)
charges (it selects the term $j=0$ in each character \chartwo\ )
\eqn\contribktwo{\eqalign{
\lim_{q \to 0} K_2(z|\tau)&=
\sum_{l=0}^k e^{4i\pi zk(k+2) ({l+1 \over k( k+2)}-{1 \over 2k})}\cr
&= e^{-2i \pi k z} {1 -e^{4i \pi (k+1)z} \over 1-e^{4i \pi z}} \cr
&= { \sin 2 \pi (k+1)z \over  \sin 2 \pi z} \cr}}
We conclude that the $q \to 0$ limit of the elliptic modular function
\sipro\ is just $1$.
In appendix A, we prove that the only elliptic modular
function of weight zero (invariant under S and T)
with a $q \to 0$ limit equal to $1$,
equals $1$ identically, hence
the poles $b_i$ cancel the zeroes $a_i$
exactly and $A=1$, and we get the desired identity
$$ K_2(z|\tau)=Z_2(2z|\tau).$$

The proof is very enlightening and suggests that somehow the elliptic
genus mainly depends on the transformations of the various fields under
the U(1) symmetry, which can be read off from just the U(1) theta
function piece of the N=2 superconformal characters.

\subsec{Generalization to D and E modular invariant theories.}

In view of the previous subsection, it is a straightforward exercise to
try to guess a product formula for say a general sum of Ramond
chiral characters, by just looking at its $q \to 0$ limit.
If the answer has the form
$$ \prod_i{ \sin 2 \pi z k_i \over \sin 2 \pi z l_i}$$
for some integers $k_i, \ l_i$, then we have to compare the
$z \to z+1$, $z \to z+ \tau$ and
$(z,\tau) \to ({z \over \tau},-{1 \over \tau})$
transformations of the sum of
characters, and of the simplest guess for the answer
$$ \prod_i {\Theta_1(2k_i z | \tau) \over \Theta_1(2 l_i z|\tau)}.$$
As expected, they will differ in general, but remarkably
when the sum over Ramond chiral characters pertains to the D and E
series of modular invariants for the $SU(2)_k$ N=2 theories,
they actually coincide, and we obtain generalizations of the identity
\witconj\ for these theories. This translates into a link with the
Landau--Ginzburg theories based on D and E type singularities
\ACT.

\noindent{\it The D case.}
The elliptic genus for a D theory at (even) level $k$ reads
\eqn\ellipgend{ K_2^{D}(z|\tau)=\sum_{l \in Exp(D)}
\chi_l^l(z|\tau) }
where $Exp(D)$ denotes the set of Coxeter exponents of the
corresponding Lie algebra $D_{{k \over 2}+2}$, shifted by one,
namely $Exp(D)=\{0,2,4,...,k,{k \over 2}\}$.
If we compare this to the previous expression for the $A$ series
\genelip, it is clear that we did not spoil the behaviour
of the function under $z \to z+1$ and $z \to z+ \tau$ \ktwoonetau.
However, this particular combination modifies the $q \to 0$
limit, which becomes
\eqn\qtoo{
\lim_{q \to 0} K_2^{D}(z|\tau)= \sum_{l \in Exp(D)}
e^{4i\pi z k(k+2)( {l+1 \over k(k+2)} -{1 \over 2k} )} }
If we define $x=e^{4i\pi z}$, this is easily seen to be
\eqn\dsi{\eqalign{x^{-{k \over 2}}
(1+x^2+x^4+...+x^k+x^{k \over 2})&=
x^{-{k \over 2}} (1+x^{k \over 2}){1-x^{{k \over 2}+2}
\over 1-x^2}\cr
&={\sin 2 \pi k z \sin  \pi (k +4)z
\over \sin \pi k z \sin 4 \pi z} \cr}}
The modular transformations are also found to be
$$\eqalign{ K_2^{D}(z|\tau+1)&=K_2^{D}(z|\tau)\cr
K_2^{D}({z \over \tau}|-{1 \over \tau})&=
e^{4i\pi k(k+2){z^2 \over \tau}} K_2^{D}(z|\tau)\cr}$$
By using the technique of previous section (the
$z \to z+1$, $z \to z+\tau$ and
$(z,\tau) \to ({z \over \tau},-{1 \over \tau})$
transformations are identical,
hence the ratio is elliptic and modular invariant
with a $q \to 0$ limit equal to $1$
and it
is therefore identically equal to one.), we find that
\eqn\gend{K_2^{D}(z|\tau) =
{\Theta_1(2kz|\tau) \Theta_1((k +4)z|\tau)
\over \Theta_1(kz|\tau) \Theta_1(4z|\tau)}.}

\noindent{\it $E_6$ case.}
The elliptic genus reads
$$K_2^{E_6}(z|\tau)=\sum_{l \in Exp(E_6)} \chi_l^l(z|\tau),$$
where $Exp(E_6)=\{0,3,4,6,7,10 \}$. With the same notations as above,
the $q \to 0$ limit reads
$$\eqalign{
x^{-5}(1+x^3+x^4+x^6+x^7+x^{10})&= x^{-5}{(1-x^8)(1-x^{9})\over
(1-x^4)(1-x^3)}\cr
&={\sin 16 \pi z \sin 18 \pi z \over \sin 8 \pi z \sin 6 \pi z}\cr
},$$
and we get
\eqn\genesix{ K_2^{E_6}(z|\tau)=
{\Theta_1(16 z|\tau) \Theta_1(18 z|\tau)
\over \Theta_1(8z|\tau) \Theta_1(6z|\tau)}.}

\noindent{\it $E_7$ case.}
Analogously,
$$K_2^{E_7}(z|\tau)=\sum_{l \in Exp(E_7)} \chi_l^l(z|\tau),$$
with $Exp(E_7)=\{0,4,6,8,10,12,16\}$, and the $q\to 0$ limit
reads
$$\eqalign{
x^{-8}(1+x^4+x^6+x^8+x^{10}+x^{12}+x^{16})&=
x^{-8}{(1-x^{12}) (1-x^{14}) \over (1-x^6)(1-x^4)} \cr
&= {\sin 24 \pi z \sin 28 \pi z \over \sin 12 \pi z \sin 8 \pi z}
\cr},$$
and we find
\eqn\geneseven{ K_2^{E_7}(z|\tau)=
{\Theta_1(24z|\tau)\Theta_1(28z|\tau)
\over \Theta_1(12z|\tau) \Theta_1(8z|\tau)} .}

\noindent{\it $E_8$ case.}
Finally,
$$K_2^{E_8}(z|\tau)=\sum_{l \in Exp(E_8)} \chi_l^l(z|\tau),$$
with $Exp(E_8)=\{0,6,10,12,16,18,22,28\}$ and the
$q \to 0 $ limit reads
$$\eqalign{ x^{-14}(1+x^6+x^{10}+x^{12}+x^{16}+x^{18}+x^{22}+x^{28})
&=x^{-14} {(1-x^{24})(1-x^{20}) \over
(1-x^6)(1-x^{10})} \cr
&={\sin 48 \pi z \sin 40 \pi z \over \sin 12 \pi z \sin 20 \pi z}
\cr},$$
and
\eqn\geneeight{ K_2^{E_8}(z|\tau)=
{\Theta_1(48z|\tau)\Theta_1(40z|\tau)\over
\Theta_1(12z|\tau) \Theta_1(20z|\tau)}.}

\subsec{N=2 Landau-Ginzburg calculation of the elliptic genus
in the D and E cases.}

We now want to match the results for the elliptic genus of the D and E
superconformal field theories against their effective Landau--Ginzburg
theories.

\noindent{\it D case.}
The potential for the $D_{{k \over 2}+2}$ theory is believed to be
\ACT\
$$W_D= { \Phi_1^{{k \over 2}+1} \over {k+2}} +\Phi_1 \Phi_2^2,$$
with $\Phi_i=\phi_i+\theta_+\psi_i^+ + \theta_- \psi_i^-+\theta_+
\theta_- F_i$, $i=1, \ 2,$ two $N=2$ superfields.
Following Witten, we identify the U(1) transformations of the lower and
fermionic components of the fields which preserve the lagrangian,
after the standard Gaussian integration over the upper components $F_i$,
as
$$\eqalign{
\phi_1 &\to e^{4i\pi u} \phi_1 \cr
\phi_2 &\to e^{i\pi k u} \phi_2 \cr
\psi_1^+ &\to e^{4i\pi u} \psi_1^+ \cr
\psi_2^{+} &\to e^{i\pi k u} \psi_2^+ \cr
\psi_1^- &\to e^{-2i \pi k u} \psi_1^- \cr
\psi_2^- &\to e^{-i\pi (k+4) u} \psi_2^- \cr.}$$
Next we perform the analogue of the A case free field computation,
which corresponds to multiplying the potential term by some
parameter $\epsilon$ and taking $\epsilon \to 0$. This does not affect
the result for the elliptic genus, due to its topological character.
Collecting the contributions of right and left movers and all the
zero and non--zero modes of the various fields,
we get the elliptic genus in infinite product form, which can be
recast thanks to the product identity \deftet\ as a simple
product of theta functions
\eqn\dpro{ Z_2^{D}(u|\tau)= {\Theta_1(ku|\tau) \Theta_1((k+4)u/2|\tau)
\over \Theta_1(ku/2|\tau) \Theta_1(2u|\tau)} }
Comparing this with the character formula \gend, we find that
$$Z_2^{D}(u=2z|\tau)=K_2^{D}(z|\tau).$$

\noindent{\it $E_6$ case.}
The potential reads \ACT\
$$W_{E_6}= {\Phi_1^4 \over 4} + {\Phi_2^3 \over 3},$$
the theory is therefore factorized into two A type theories,
so is the elliptic genus.
Taking into account the quasi--homogeneity degree of the potential
(12 here) which provides us with a link
between the U(1) charges in the two A theories,
we find the following transformations for the field
components
$$\eqalign{
\phi_1 &\to e^{6i\pi u} \phi_1 \cr
\phi_2 &\to e^{8i\pi  u} \phi_2 \cr
\psi_1^+ &\to e^{6i\pi u} \psi_1^+ \cr
\psi_2^{+} &\to e^{8i\pi  u} \psi_2^+ \cr
\psi_1^- &\to e^{-18i \pi  u} \psi_1^- \cr
\psi_2^- &\to e^{-16i\pi u} \psi_2^- \cr.}$$
and the elliptic genus reads finally
$$Z_2^{E_6}(u|\tau)=
{\Theta_1(9u|\tau) \Theta_1(8u|\tau)
\over \Theta_1(4u|\tau) \Theta_1(3u|\tau)},$$
which coincides with \genesix\ for $u=2z.$

\noindent{\it $E_7$ case.}
The potential reads \ACT\
$$W_{E_7}= {\Phi_1^3 \over 3} + \Phi_1 \Phi_2^3,$$
and we have the following transformations for the field components
$$\eqalign{
\phi_1 &\to e^{12i\pi u} \phi_1 \cr
\phi_2 &\to e^{8i\pi  u} \phi_2 \cr
\psi_1^+ &\to e^{12i\pi u} \psi_1^+ \cr
\psi_2^{+} &\to e^{8i\pi  u} \psi_2^+ \cr
\psi_1^- &\to e^{-24i \pi  u} \psi_1^- \cr
\psi_2^- &\to e^{-28i\pi u} \psi_2^- \cr.}$$
and the elliptic genus reads finally
$$Z_2^{E_7}(u|\tau)=
{\Theta_1(12u|\tau) \Theta_1(14u|\tau)
\over \Theta_1(6u|\tau) \Theta_1(4u|\tau)},$$
identical to \geneseven\ up to $u=2z$.

\noindent{\it $E_8$ case.}
The potential reads \ACT\
$$W_{E_8}= {\Phi_1^5 \over 5} + {\Phi_2^3 \over 3},$$
and the theory factorizes again into two A type theories,
and so does the elliptic genus.
We have the following U(1) transformations of the fields
$$\eqalign{
\phi_1 &\to e^{12i\pi u} \phi_1 \cr
\phi_2 &\to e^{20i\pi  u} \phi_2 \cr
\psi_1^+ &\to e^{12i\pi u} \psi_1^+ \cr
\psi_2^{+} &\to e^{20i\pi  u} \psi_2^+ \cr
\psi_1^- &\to e^{-40i \pi  u} \psi_1^- \cr
\psi_2^- &\to e^{-48i\pi u} \psi_2^- \cr.}$$
and the elliptic genus reads finally
$$Z_2^{E_8}(u|\tau)=
{\Theta_1(24u|\tau) \Theta_1(20u|\tau)
\over \Theta_1(6u|\tau) \Theta_1(10u|\tau)},$$
identical to \geneeight\ up to $u=2z$.

This completes the identification of elliptic genera for
D and E Landau--Ginzburg potentials and that of the corresponding
superconformal field theories.

\newsec{The SU(N) case.}

We will consider now a SU(N) generalization of the above SU(2) $N=2$
superconformal theories introduced by Kazama and Suzuki
\KS. The theory is best expressed as a coset
of the form
\eqn\cosetN{\eqalign{
{SU(N)_k \times SO(2(N-1))_1 \over SU(N-1)_{k+1} \times U(1)}&\equiv
\left[{SU(N)_k \over SU(N-1)_{k}\times U(1)}\right] \times \cr
&\ \ \ \ \times \left[{SU(N-1)_k\times
SU(N-1)_1 \over SU(N-1)_{k+1}}\right] \times \big[U(1)\big] ,\cr}}
where we used the conformal embedding $SO(2(N-1))_1\to
SU(N-1)_1 \times U(1)$, expressing the theory of $2(N-1)$
Majorana fermions as that of $N-1$ Dirac fermions with a
U(1) symmetry generated by some bilinear in the original fermions.
Clearly the $N=2$ characters will decompose into three pieces
\ref\YN{D. Nemeschansky, K. Huitu and S. Yankielowicz,
Phys. Lett. {\bf B246} (1990) 105.},
pertaining to the three expressions between brackets in eqn.\cosetN,
with respective branching functions $b$, minimal characters $\chi$
and U(1) characters $\Theta$.
For the Ramond sector, they read
\eqn\charN{\eqalign{
\chi_{\lambda^{(N-1)},q}^{\lambda^{(N)}}(z|\tau)&=
\sum_{m,\mu^{(N-1)}} \ b_{\mu^{(N-1)},m}^{\lambda^{(N)}}(\tau)
\times \cr
&\times \chi_{\mu^{(N-1)},\lambda^{(N-1)}}(\tau)\
\Theta_{(k+N)(m-q-\sigma)+N(q+\sigma),N^2(N-1)k(k+N)/2}(z|\tau),\cr}}
where $\lambda^{(M)}=\sum_{i=1}^{M-1} \lambda_i^{(M)} \omega_i^{(M)}$,
($\omega_i^{(M)}$ the fundamental weights of SU(M))
denotes an integrable weight of SU(M) at the corresponding level $p$,
i.e. subject to $\lambda_i^{(M)} \geq 0$ and $\sum_i \lambda_i^{(M)}
\leq p$ (in the following we will use the
notation $P_p^{(M)}$ for the set
of allowed $\lambda_i^{(M)}$.).
We choose the convention that the U(1) charge $q$ be an integer, and
it gets shifted by an integer $\sigma$ in the Ramond sector. Finally
the U(1) character is given by a theta function as defined in
eqn.\defthet.
For the computation of the elliptic genus of the $\cal A$ type
models, we only need a sum
over the Ramond states with $L_0-c/24=0$, which amounts to
\RING\ \EMLG\
\ref\GE{D. Gepner, Comm. Math. Phys. {\bf 142} (1991) 433.}\
$$\eqalign{
\lambda_i^{(N-1)} &=\lambda_i^{(N)} \ \ \ i=1,2,...,N-2 \cr
q&= \sum_{i=1}^{N-1} i \lambda_i^{(N)} \cr
\sigma &= N(N-1)/2 \cr}$$
So the final expression for the elliptic genus of the N=2
Kazama--Suzuki theories reads
\eqn\ellipgeN{ K_N(z|\tau)=\sum_{\lambda^{(N)} \in P_k^{(N)}}
{\chi}_{\lambda^{(N-1)},q}^{\lambda^{(N)}}(z|\tau). }

\subsec{A product formula for the SU(N) elliptic genus.}

Following the lines of the SU(2) proof of sect.2, we wish to
study the $z \to z+1$, $z \to z+\tau$ and
$(z,\tau)\to ({z \over \tau},-{1 \over \tau})$ transforms of the
elliptic genus $K_N$, together with its $q \to 0$ limit
(as noted before, the $\tau \to \tau+1$ invariance is built--in
in the definition of the elliptic genus).
The latter will suggest a product formula for $K_N$, which we
will eventually prove by elliptic modular function techniques,
using the lemma of appendix A.
It is straightforward to see that
\eqn\traN{\eqalign{ K_N(z+1|\tau)&=K_N(z|\tau) \cr
K_N(z+\tau|\tau)&=
e^{-i\pi N^2 (N-1)k(k+N)(\tau+2z)} K_N(z|\tau).\cr}}
The modular transformations of the Ramond characters are
cumbersome \GE, and we omit their details here, they lead to the
final covariance property of the elliptic genus under the S
transformation
$$K_N({z \over \tau}|-{1 \over \tau})=e^{i\pi N^2(N-1)k(k+N)
{z^2 \over \tau}} K_N(z|\tau).$$
Finally, the $q \to 0$ limit is again entirely given by
the U(1) charges of the Ramond fields (it corresponds to
only the terms with $\mu^{(N-1)}=\lambda^{(N-1)}$, and $m=q$
in the sum \ellipgeN\ with \charN\ ), as
\eqn\limN{ \sum_{\Gl_1,...,\Gl_{N-1} \in P_k^{(N)}}
e^{2i\pi zN(\Gl_1+2\Gl_2+...+(N-1)\Gl_{N-1} -k(N-1)/2)} }
Upon introducing the variable $x=e^{2i\pi Nz}$, we get\foot{
The proof of this identity is given in appendix B below.}
$$\eqalign{ \lim_{q \to 0} K_N(z|\tau)&= x^{-k(N-1)/2}
\sum_{\Gl_i \geq 0 \atop \Sigma \Gl_i \leq k}
x^{\Sigma i \Gl_i}\cr
&= x^{-k(N-1)/2} \prod_{j=1}^{N-1} {1-x^{k+j} \over
1-x^j} \cr
&=\prod_{j=1}^{N-1} {\sin \pi N(k+j)z \over \sin \pi Nj z}\cr}$$
Using the same technique as in sect.2, and the lemma of
appendix A, it is easy to
establish the following product formula
\eqn\proN{K_N(z|\tau)= \prod_{j=1}^{N-1} {\Theta_1(N(k+j)z|\tau)
\over \Theta_1(Njz|\tau)}.}

\subsec{N=2 Landau-Ginzburg calculation of the elliptic genus for
SU(N) theories.}

We perform now the computation of the elliptic genus defined
in eqn.\ellipgen\ denoted now
$Z_N(u|\tau)$, $u=\gamma_L / 2 \pi (k+N)$, within the Landau--Ginzburg
framework.
The effective Landau--Ginzburg description of the Kazama--Suzuki
theories based on $SU(N)_k$
is believed to be the following. It is an effective theory of
$N-1$ superfields $\Phi_1,...,\Phi_{N-1}$, with components
$\Phi_i=\phi_i+\theta_+ \psi_i^++\theta_-\psi_i^- +\theta_+ \theta_-
F_i$, governed by the superpotential
$W_N^{(k+N)}(\Phi_1,...,\Phi_{N-1})$,
given in compact form by the generating function
\EMLG\
$$ \sum_{m \geq 0} t^m W_N^{(m)}(x_1,...,x_{N-1})= -\log( 1-tx_1+t^2 x_2
-t^3 x_3+...+ (-1)^{N-1}t^{N-1}x_{N-1}).$$
The important fact is that the potential $W_N^{(k+N)}$ is
quasi--homogeneous in the fields, with total degree $k+N$, for a
gradation which assigns the weight $i$ to the field $\Phi_i$,
$i=1,2,...,N-1$.
Repeating Witten's steps in the direct Landau--Ginzburg calculation of
the elliptic genus, we find the following U(1) transformations of the
field components preserving the Lagrangian of the theory
$$\eqalign{ \phi_m &\to e^{2i\pi m u} \phi_m \cr
\psi_m^+ &\to e^{2i\pi m u } \psi_m^+ \cr
\psi_m^- &\to e^{-2i\pi (k+m)u} \psi_m^- \cr}.$$
We again eliminate the potential by a
scaling factor which we send
to zero, but the result is unchanged by topological invariance of the
elliptic genus. The free field calculation is tedious but
straightforward. Putting all contributions from right and left moving
fermions and bosons, we end up with
\eqn\geN{ Z_N(u|\tau)= \prod_{j=1}^{N-1} {\Theta_1((k+j)u|\tau) \over
\Theta_1(ju|\tau)} }
This is nothing but the $N=2$ superconformal coset genus, expressed
through the product formula
\proN, with $u=Nz$
$$ Z_N(u=Nz|\tau)=K_N(z|\tau).$$

\newsec{Discussion and comments.}

The identification of the elliptic genus in both superconformal and
Landau--Ginzburg frameworks is a non--negligible piece of evidence
toward identification of the theories.
One might wonder how much the elliptic genus says
about either theory. As explicitly observed in the above computation,
the main constraint (although not sufficient) comes from the $q \to 0$
limit of the elliptic genus of the superconformal theory.
It turns out that this carries mainly the information of U(1)
charges of the primaries of the Ramond sector, with $L_0=c/24$.
On the other hand, the Landau--Ginzburg computation shows that
the elliptic genus only knows about U(1) transformations ("R--symmetry")
of the components of the basic superfields.
Previous evidence gathered so far as to identification of the
theories concerned in particular the identification between
the "chiral ring" of the superconformal theory and the polynomial
ring associated to the Landau--Ginzburg potential
${\cal R} =\IC[x_1,...,x_p]/\{ \partial_{x_i}W\}$ \EMLG\ \CHF.
The "chiral ring"
is just the set of primaries of the Neveu--Schwarz sector of the
superconformal theory
which go over to the Ramond states forming the elliptic genus under
spectral flow.
{}From our calculations, we learn that the identification goes beyond
the ring structures, but also concerns U(1) grading of ring bases.
The fact that not just a ring but a ring together with a graded basis
is the essential information needed for Landau--Ginzburg description
of say fusion rings of Rational Conformal Theories was pointed out in
\ref\JBPDF{P. Di Francesco and J.-B. Zuber,
{\it Fusion potentials I}, to appear in J. Phys {\bf A} (1993).}.

Now, having computed the elliptic genus for a given superconformal
theory, and hopefully obtained it in "product type" form, we
certainly learn something about the Landau--Ginzburg potential
describing it, if there exists any. We would like to stress here
that this might be the most powerful tool up to now for hunting
Landau--Ginzburg potentials for other N=2 superconformal theories.

\subsec{ Uniqueness in the Landau--Ginzburg potential identification.}

One might wonder in which sense the answer we found for possible
LG  descriptions of N=2 superconformal theories is unique.
Here is an example of different superconformal theories sharing the
same elliptic genus.
If we take the expression \proN\ of the $SU(N)_k$ elliptic
genus for level $k=1$, we find lots of cancellations in the
product of theta functions, so that we are left with
$$ {\Theta_1(N^2z|\tau) \over \Theta_1(Nz|\tau)}.$$
Up to a redefinition $z \to z / N$, this is nothing but the
elliptic genus of the $SU(2)_{N-1}$ superconformal theory
as expressed through eqn.\witrevis.
So we get a simple example where although the elliptic genera coincide,
the superconformal theories are different.

On the other hand, from the Landau--Ginzburg point of view, it is
easy to see that the $SU(N)_1$ fusion ring can be viewed as
a {\it perturbation} of the $SU(2)_{N-1}$ Landau--Ginzburg potential,
by adding the most relevant perturbation by $\Phi_1$:
$$ W= {\Phi_1^{N+1} \over {N+1}} -t_{N} \Phi_1, $$
where $t_N$ is a dimensionful coupling preserving the global
quasi--homogeneity of the potential.
This perturbation is known to correspond to an integrable perturbation
of the associated superconformal theory
\ref\FMVW{P. Fendley, S. Mathur, C. Vafa and N. Warner, Phys.
Lett. {\bf B243} (1990) 257; T. Eguchi and S.-K. Yang, Mod. Phys.
Lett. {\bf A5} (1990) 1693;
P. Mathieu and M. Walton, Phys. Lett. {\bf B254} (1991) 106;
P. Fendley, W. Lerche, S. Mathur and N. Warner,
Nucl. Phys. {\bf B348} (1991) 66.}.
It would seem that
the elliptic genus is indeed preserved under certain  "massive"
perturbations of the initial Landau--Ginzburg theory \WIT.
This point should certainly be the object of further study.

We find more coincidences between various elliptic genera by comparing
the $SU(N)_k$ result \proN\ to that of $SU(k+1)_{N-1}$, when
$k+1<N$. Due to cancellations in the numerator and denominator of
\proN, we end up with the same result. We believe that this corresponds
to some generalization of the above phenomenon,
that certain perturbations preserve the elliptic genus.
This is also related to the "level--rank"
duality of affine Lie algebras
\ref\ABS{D. Altsch\"uler, M. Bauer and H. Saleur,
J. Phys. Lett. {\bf A23} (1990) 789.}\ \GE\
\ref\GG{D. Gepner, {\it Foundations of Rational Conformal Field Theory I},
Caltech preprint CALT-68-1825 (1992).}.

\subsec{Some examples which do not work.}

There are still many puzzles left in the attempts to describe the
known superconformal theories in Landau--Ginzburg terms.
For instance we did not find any candidates for the potential
associated to the "$\cal D$ series" of SU(3), obtained by $\IZ_3$
orbifold of the $\cal A$ solutions
\ref\RA{P. Christe and F. Ravanini,
Int. J. Mod. Phys. {\bf A4} (1989) 897;D. Altsch\"uler,
J. Lacki and P. Zaugg, Phys. Lett. {\bf B205} (1988) 281.}.
The main reason is that the
natural grading inherited from that of $SU(3)$ primaries
(degree $\Gl_1+2 \Gl_2$ for the $(\Gl_1,\Gl_2)$ primary)
does not allow for a nice product formula for the
$q \to 0$ limit of the elliptic genus.
For instance, in the case of level $3$, the orbifold elliptic
genus has the limit
$$\lim_{q \to 0} K_3^{{\cal D}^{(3)}}(z|\tau)=
x^{-3}(1+4x^3+x^6),$$
where as usual we set $x=e^{6i\pi z}$. It is clear that this expression
will never be put into a single product of terms of the form
$(1-x^{a})^{\pm1}$(otherwise the zeroes of the
polynomial would all be
of modulus one,
which is not the case.)\foot{Actually we find many
close factorization formulae up to a constant, as
$2+x^{-3}((1-x^6)/(1-x^3))^2$,
$3+x^{-3}(1-x^9)/(1-x^3)$, or $4+x^{-3}(1-x^{12})/(1-x^6)$,
and this is general for $\cal D$ SU(3) cases,
although we still do not understand how
to use this property efficiently.}.
This may be an indication that the grading we took for granted is
actually wrong, emphasizing again the importance of the choice of a
graded basis of the chiral ring.
It turns out actually that in this particular example
\ref\PJN{P. Di Francesco, F. Lesage, N. Sochen
and J.-B. Zuber, papers in preparation.},
the (dual) ring
associated with the generalized Dynkin diagram
${\cal D}^{(3)}$
of $SU(3)_3$ can be seen as a particular perturbation of the
$SU(2)_{8}$--$D_6$ Landau--Ginzburg polynomial ring. But in this
picture, the natural grading of the ring is that of $D_6$,
i.e. fields with degrees $\in \{0,2,4,4,6,8\}$ (instead of
$\{ 0,3,3,3,3,6\}$, inherited from SU(3)).
It is tempting to think that in general the correct grading for
$\cal D$ type SU(3) theories will not be that inherited from the
SU(3) primaries, but one which restores the property of factorizability
of the $q \to 0$ limit of the elliptic genus.
But a different grading also means a different definition of
the U(1) charge from that of the N=2 superconformal algebra, which
sounds very bizarre.
If on the contrary we believe the initial grading is correct, there
might be something wrong with the original free field computation
of the LG elliptic genus, for instance that the R--symmetry is not
well diagonalized on the fields forming the hypothetic LG potential.

\subsec{A criterium for non--existence of a LG description?}

By reversing the argument, this may also be an indication that no LG
description exists for these cases.
An interesting consequence of the existence of a polynomial LG potential
for a graded ring is that it is a  polynomial ring of say $p$ variables
$x_1,...,x_p$,
with exactly $p$ constraints in the form of $\partial_{x_i}W=0$, and with
{\it no relation} between the constraints.
Therefore the generating function for degrees of the elements of the
ring takes the form
\eqn\genlg{ \sum_{{\rm ring}\ {\rm elements}} x^{\rm degree}=
\prod_{i=1}^p {1-x^{R_i} \over 1-x^{r_i}},}
where $r_i$ denotes the degree of the field $x_i$ and $R_i$ the degree
of the $i^{th}$ constraint $\partial_{x_i}W$, hence
$R_i=m-r_i$, if $m$ denotes the total degree of the potential $W$.
In general, the elliptic genus is a modular form generalizing this function.
So whenever the U(1) grading of the Ramond states entering
the definition of the elliptic genus does not yield a generating
function of the form \genlg,
we may conclude that there is no Landau--Ginzburg
description of the theory, at least none with the purported
grading of chiral fields.

But even if this condition is fulfilled, it is possible to find examples
where the $q \to 0$ limit of the elliptic genus is nicely factorized,
although the elliptic genus itself is not, signaling again that
something goes wrong with the hypothetic LG description.
A first example of this kind is the projection of the elliptic genus
onto a (closed) subset of Ramond states.
Take for instance the $SU(2)$ case
of sect.2, and consider the projected sum
\eqn\redsu{ \sum_{l=0,k} \chi_l^l(z|\tau) }
The $q \to 0$ limit of this sum is just (we set $x=e^{4i\pi z}$)
$$ x^{-k/2}(1+x^k)= x^{-k/2} {1-x^{2k} \over 1-x^k}={\sin 4 \pi k z
\over \sin 2 \pi k z},$$
which obviously does not lead to the analogous ratio of $\Theta_1$
for the projected sum \redsu, due to difference
between the $z \to z+\tau$
transformations.
It seems that although such a projection makes sense at the level of fusion
rings (this corresponds to taking the subring of $\{1,x,x^2,...,x^k\}$,
$x^{k+1}=0$,
formed by $\{ 1, x^k\}$, $x^{2k}=0$.), it violates higher loop consistency
(e.g. modular invariance, or the fact that the elliptic genus is a modular
form of weight zero).

The second example we wish to study is of a slightly different nature,
as it is not expected to violate modular invariance.
Consider the SU(3) Kazama--Suzuki theory with modular invariant
arising from the $SU(3)_{9}$ exceptional modular invariant
(see for instance \RA, and
\ref\TG{P. Ruelle, E Thiran and J. Weyers,
{\it Implications of an arithmetical
symmetry of the commutant for modular invariants}, DIAS preprint
STP--92--26 (1992); T. Gannon,
{\it The Classification of SU(3) modular
invariant partition functions}, Carleton preprint
92-061 (1992);
T. Gannon and Q. Ho-Kim, {\it The low level modular invariant
partition functions of rank two algebras}, Carleton
preprint 93-0364 (1993).}\
for a complete classification)
which reads
\eqn\charenine{
{\cal E}^{(9)}=|\chi_{0,0}+\chi_{9,0}+\chi_{0,9}|^2
+|\chi_{2,2}+\chi_{5,2}+\chi_{2,5}|^2+|\chi_{3,0}+
\chi_{6,3}+\chi_{0,6}|^2+|\chi_{0,3}+\chi_{3,6}+\chi_{6,0}|^2}
where the notation $(\Gl_1,\Gl_2)$ stands for an integrable
weight of SU(3) at level $9$, i.e. an element of $P_9^{(3)}$.
The elliptic genus for this theory is expressed in terms of
the Ramond characters of
$SU(3)_9\times SO(4)_1/SU(2)_{10}\times U(1)$
$$K_3^{{\cal E}^{(9)}}(z|\tau)=\sum_{(\Gl_1,\Gl_2)\in
Exp({\cal E}^{(9)})}
\chi_{\Gl_1,\Gl_1+2\Gl_2}^{(\Gl_1,\Gl_2)}(z|\tau)$$
where $Exp({\cal E}^{(9)})$ is the set of all the
couples appearing in eqn.\charenine, generalizing the set
of Coxeter exponents of the SU(2) case
(see \ref\PJB{P. Di Francesco
and J.-B. Zuber, Nucl. Phys. {\bf B338} (1990) 602;
P. Di Francesco, Int. J. Mod. Phys. {\bf A7},
No.3 (1992) 407.} for the associated
generalization of Dynkin diagrams, graph rings, etc...).
Again, the $z \to z+1$ and $z \to z+\tau$ transformations
are the same as in the $SU(3)$ case at level $9$, the only modification
affects the $q \to 0$ limit of the elliptic genus, which now reads
$$\eqalign{ \lim_{q \to 0} K_3^{{\cal E}^{(9)}}(z|\tau)&=
x^{-9}(1+x^3+3x^6+2x^9+3x^{12}+x^{15}+x^{18})\cr
&= x^{-9}{1-x^9 \over 1-x^3}\left({1-x^{12} \over 1-x^6}
\right)^2 \cr
&= {\sin 27 \pi z \over \sin 9 \pi z }\left({\sin 36 \pi z
\over \sin 18 \pi z}\right)^2,\cr}$$
where, as before, we use the variable $x=e^{6i\pi z}$.
The problem here is that the ratio of products of theta functions
$$ {\Theta_1(27z|\tau)
\over \Theta_1(9z|\tau)}\left({\Theta_1(36z|\tau) \over
\Theta_1(18z|\tau)}\right)^2$$
does not have the same $z \to z+\tau$ transformation
as the elliptic genus
$K_3^{{\cal E}^{(9)}}(z|\tau)$, and therefore we have no product
formula for the result.
Before making this remark we would have naively
learnt from this expression that if a potential description
exists, it might correspond to the tensor product of an $A$ type
SU(2) potential at level $2$ (first ratio of sines)
by a $D_4$ potential of SU(2) at level $4$ (square ratio of sines,
compare with \dsi\ with $k=4$),
i.e. with potential
$${\Phi_1^4 \over 4} +
{\Phi_2^3 \over 3} -\Phi_2 \Phi_3^2. $$
The only problem is that this is not quasi--homogeneous, as
$\Phi_1$ should have degree $3$, and $\Phi_2$ and $\Phi_3$ degree $6$.
So we run into inconsistencies in our search for a potential for
${\cal E}^{(9)}$, and the U(1) grading is probably such
that no LG description of this theory exists at all.

Many more theories do not pass the test of eqn.\genlg,
let us just quote the case of $G(2)$ models\ref\MBP{
M. Bauer and P. Di Francesco, in progress.}, for which the grading of
Ramond states with $L_0=c/24$ violates \genlg. It would seem that
actually only $SU(N)$ and $SP(N)$ models, for which LG potentials are
known
\EMLG\
\ref\GS{D. Gepner and A. Schwimmer, Nucl. Phys. {\bf B380}
(1992) 147.}, pass the test successfully.

\bigskip
\bigskip

\noindent{\bf Acknowledgements}

\noindent{}We would like to thank O. Aharony for a
useful discussion and especially J.-B. Zuber
for a careful and critical reading of the manuscript.
P. D.F. would like to thank the members of
the theory group at
the School of Physics and Astronomy, Tel--Aviv University, for their
warm hospitality and stimulating discussions, while most of
this work was
completed.

\bigskip
\appendix{A}{A useful lemma on elliptic modular functions.}

We wish to prove the following lemma.

\noindent{}Let $f(z|\tau)$ be an elliptic modular
function, subject to
$$\eqalign{(i) \ \ \ f(z+1|\tau)&=f(z|\tau)\cr
(ii) \ \ \ f(z+\tau|\tau)&=f(z|\tau)\cr
(iii)\ f({z \over \tau}|-{1 \over \tau})&=f(z|\tau)\cr
(iv)\ \ \ f(z|\tau+1)&=f(z|\tau),\cr}$$
and such that
$$lim_{\tau \to i \infty} f(z|\tau) =1.$$
Then $f$ is identically equal to $1$.

$f$ being elliptic, it has the form \WWAT\
\eqn\prof{f(z|\tau)=A
\prod_{i=1}^n {\Theta_1(z-a_i(\tau)|\tau)\over
\Theta_1(z-b_i(\tau)|\tau)}.}
Let us proceed and show that the dependence of the
zeroes and poles on the modular parameter
$\tau$ is linear, thanks to the "S" and "T"
invariance (iii) and (iv).
The S invariance (iii) implies that there exists a permutation
$\sigma$ of $\{ 1,2,...,N\}$ such that
\eqn\Sa{ \tau a_i(-{1 \over \tau})=a_{\sigma(i)}(\tau)+m_i+n_i\tau }
(the zeroes are just permuted under the transformation, up to
integer shifts $m_i$ and $n_i$ of $1$ and $\tau$.).
Analogously, the T invariance implies that there exists a permutation
$\rho$ such that
$$a_i(\tau+1)=a_{\rho(i)}(\tau)+p_i+q_i\tau,$$
for some integers $p_i$ and $q_i$, hence
\eqn\Ta{a_i(\tau+K)=a_i(\tau)+ r_i+s_i\tau,}
where $K$ denotes the order of the permutation $\rho$,
and $r_i$, $s_i$ are some integers.
Iterating this $l$ times, we get
$$a_i(\tau+Kl)=a_i(\tau)+l r_i +{l(l-1)\over 2} s_i +l s_i \tau.$$
Combining this relation for $i \to \sigma(i)$ with \Sa, we have
$$(\tau+Kl)a_i(-{1 \over \tau+Kl}) -\tau a_i(-{1 \over \tau})=
lr_{\sigma(i)}+{l(l-1)\over 2}s_{\sigma(i)}+l(Kn_i+s_{\sigma(i)})\tau.$$
The zeroes are analytic functions of the modular parameter $\tau$,
therefore we can compute the large $l$ expansion of the above, which
imposes that $s_{\sigma(i)}=0$ ($l^2$ term),
$K a_i(0)=r_{\sigma(i)}$ and $n_i=0$ ($l$ term),
and the constant term yields
$$\tau a_i(-{1 \over \tau})= a_i(0)\tau -a_i'(0),$$
which means that the zeroes are linear functions of $\tau$
(and so are the poles, but those are under control in the formulas
we establish in this paper, as they arise from the known zeroes of the LG
elliptic genera $Z$.).
{}From eqn.\Ta, we also find that they have the general form
$$a_i(\tau)={1 \over L}(\alpha_i+\beta_i \tau),$$
where we take for $L$ the smallest common multiple of all the $K$'s,
when we run over the index $i$, and $\alpha_i,\beta_i$ are integers
in $\{0,1,...,L-1\}$.
We denote by $(\alpha_i,\beta_i)$ the corresponding zero.
Under the S and T transformations, the zeroes become, up to
sums of integer multiples of $L$ and $L\tau$
$$\eqalign{ S\ : \ \ (\alpha,\beta) &\to (-\beta,\alpha) \cr
T\ : \ \ (\alpha,\beta) &\to (\alpha+\beta,\beta).\cr}$$
This is nothing but the action of the modular transformations
on the $a$ and $b$ homology cycles of a torus. In this framework,
it is known that for any couple $(\alpha,\beta)$ of winding
numbers, there exists a modular transformation
$\phi=\prod S^{m_i}T^{n_i}$, such that
$$\phi(\alpha,\beta)=(\alpha \wedge \beta,0),$$
where $\alpha \wedge \beta$ denotes the greatest common divisor
of $\alpha$ and $\beta$.
Hence the modular transformation $\phi$ sends $a_i(\tau)$
to a real zero (with no $\tau$ component).

But the real zeroes of $f$ are constrained by the $q \to 0$ limit
of $f$, which becomes, by a straightforward use of the definition
of $\Theta_1$
$$1=A \prod_{a,b,\  {\rm real}} {\sin \pi(z-a) \over \sin \pi(z-b)}
\lim_{q \to 0}
\prod_{c,d \ {\rm complex}} e^{-i \pi (c-d)} q^{-(c-d)/2}, $$
where we distinguished between the real zeroes ($a$) and poles
($b$), and those (resp. $c$ and $d$) with a non zero $\tau$ component.
This implies that all the {\it real} zeroes are exactly cancelled by
{\it real} poles, and one gets $A=1$, in addition to some usual
sum rule for the other zeroes and poles.
Now, as we showed above, any {\it non--real} zero can be
transformed into a {\it real} one by some modular transformation $\phi$,
which leaves $f$ unchanged, thanks to (iii) and (iv).
But we just saw that no real zero of $f$ can survive, as it has to be
cancelled by a pole.
Therefore no zero {\it at all} can survive in the product \prof,
and the function $f$ is identically equal to $1$.

\appendix{B}{Proof of the SU(N) product formula.}

We wish to prove that
$$A_k^{(N)}(x)=\sum_{\Gl_1,...,\Gl_{N-1} \in P_k^{(N)}}
x^{\Sigma i\Gl_i}
=\prod_{j=1}^{N-1} {1 - x^{k+j} \over 1-x^j}.$$
Changing variables in the summation to
$l_i=\Gl_i+\Gl_{i+1}+...+\Gl_{N-1}$, we get
$$\eqalign{
A_k^{(N)}(x)&=
\sum_{0 \leq l_{N-1}\leq l_{N-2}\leq ...\leq l_1 \leq k}
x^{l_1+l_2+...+l_{N-1}} \cr
&=\sum_{l=0}^{k} x^l A_l^{(N-1)}(x). \cr}$$
The property above is clear for $N=2$ and any $k$.
Let us now proceed by recursion: suppose the property is proved
for any $A_m^{(P)}$, with $m+P \leq N+k-1$, let us prove it for
$A_k^{(N)}$.
Thanks to the recursion hypothesis, we have
$$A_k^{(N)}(x)= \sum_{l=0}^k x^l \prod_{j=1}^{N-2}
{1-x^{l+j} \over 1-x^j}.$$
Introduce $B_k^{(N)}(x)= A_k^{(N)}(x) \prod_{j=1}^{N-1}
(1-x^j)$, then we have
$$B_k^{(N)}(x)=(1-x^{N-1}) \sum_{l=0}^k x^l \prod_{j=1}^{N-2}
(1-x^{l+j}).$$
Now expand the product as
$$ \prod_{j=1}^{N-2} (1-x^{l+j})= \sum_{r=0}^{N-2} (-x)^{lr}
\sigma_r^{(N-1)}(x), $$
where
$$\sigma_r^{(N-1)}(x)= \sum_{1 \leq j_1 < ... < j_r \leq N-2}
(-x)^{j_1+...+j_r}. $$
Performing the sum over $l$, we get (after a shift $r \to r+1$
of the summation variable)
$$ B_k^{(N)}(x)= (1-x^{N-1})
\sum_{r=1}^{N-1} {1-(-1)^{(r-1)(k+1)}x^{r(k+1)}
\over 1-(-1)^{r-1}x^{r}}
\sigma_{r-1}^{(N-1)}(x).$$
Now rewrite the term
$${1-(-1)^{(r-1)(k+1)}x^{r(k+1)}
\over 1-(-1)^{r-1}x^r}={1-(-1)^{k(r-1)}x^{rk}
\over 1-(-1)^{r-1}x^r}+ (-1)^{k(r-1)}x^{rk},$$
hence
$$\eqalign{
B_k^{(N)}(x)&=B_{k-1}^{(N)}(x)+(1-x^{N-1})x^k
\prod_{j=1}^{N-2}(1-x^{k+j})\cr
&=[1-x^k +(1-x^{N-1})x^k] \prod_{j=1}^{N-2} (1-x^{k+j})\cr
&=\prod_{j=1}^{N-1}(1-x^{k+j}),\cr}$$
where we used the recursion hypothesis to express $B_{k-1}^{(N)}$.
Therefore we get
$$A_k^{(N)}(x)=\prod_{j=1}^{N-1} {1-x^{k+j} \over 1-x^j}.$$

\listrefs
\end